\documentclass[12pt]{article}
\usepackage{amsmath}
\usepackage{graphicx,psfrag,epsf}
\usepackage{enumerate}
\usepackage{natbib}
\usepackage{url} 

\usepackage{amsfonts}
\usepackage{amssymb}
\usepackage{floatrow}
\usepackage{float}
\usepackage{bm}
\usepackage{ulem}
\usepackage[title]{appendix}

\newcommand{\blind}{1}

\begin{document}

\def\spacingset#1{\renewcommand{\baselinestretch}%
{#1}\small\normalsize} \spacingset{1}


\if1\blind
{
  \title{\bf{A Hierarchical Framework for Correcting Under-Reporting in Count Data}}
  \author{Oliver Stoner\thanks{
    The authors gratefully acknowledge the funding of this research in part by the Engineering and Physical Sciences Research Council (EPSRC) and in part by the Nature and Environment Research Council (NERC). The associate editor and two anonymous reviewers also contributed substantially to making this article more comprehensive.}\hspace{.2cm}\\
    Department of Mathematics, University of Exeter, Exeter, UK\\
    and\\
    Theo Economou \\
    Department of Mathematics, University of Exeter, Exeter, UK\\
    and\\
    Gabriela Drummond \\
    University of Bras\'{i}­lia, Bras\'{i}­lia, Brazil}
  \maketitle
} \fi

\if0\blind{
 \title{\bf{A Hierarchical Framework for Correcting Under-Reporting in Count Data}}
\maketitle

} \fi

\bigskip
\begin{abstract}
Tuberculosis poses a global health risk and Brazil is among the top twenty countries by absolute mortality. However, this epidemiological burden is masked by under-reporting, which impairs planning for effective intervention. We present a comprehensive investigation and application of a Bayesian hierarchical approach to modelling and correcting under-reporting in tuberculosis counts, a general problem arising in observational count data. The framework is applicable to fully under-reported data, relying only on an informative prior distribution for the mean reporting rate to supplement the partial information in the data. Covariates are used to inform both the true count generating process and the under-reporting mechanism, while also allowing for complex spatio-temporal structures. We present several sensitivity analyses based on simulation experiments to aid the elicitation of the prior distribution for the mean reporting rate and decisions relating to the inclusion of covariates. Both prior and posterior predictive model checking are presented, as well as a critical evaluation of the approach.

\end{abstract}

\noindent%
\vfill

\newpage
\spacingset{1.45} 

\section{Introduction}
In a variety of fields, such as epidemiology and natural hazards, count data arise which may not be a full representation of the quantity of interest. In many cases the counts are under-reported: the recorded value is less than the true value, sometimes substantially. Quite often, this is due to the observation process being flawed, for instance failing to reach some individuals in a population at risk from infectious disease such as tuberculosis or TB, which is the motivating application here. It is then a missing data challenge and from a statistical point of view, a prediction problem.

The TB surveillance system in Brazil is responsible for detecting disease occurrence and for providing information about its patterns and trends. The notification of TB is mandatory and the data are available in the Notifiable Diseases Information System (SINAN), which provides information about the disease at national, state, municipal and other regional levels. Despite the high spatial coverage of SINAN, the system is not able to report all TB cases. Using inventory studies \citep{TBInventory}, the overall TB detection rate for Brazil was estimated as 91\%, 84\%, and 87\% for the years 2012 to 2014 \citep{TBReport}.

Under-reporting is an issue because it can lead to biased statistical inference, and therefore poorly informed decisions. This bias will affect parameter estimates, predictions and associated uncertainty. Conventional approaches to quantifying risk, for instance by estimating the spatio-temporal disease rate per unit population, are liable to under-estimate the risk if under-reporting is not allowed for. This has serious societal implications---an estimated 7300 deaths were caused by TB in Brazil in 2016 \citep{TBReport}, and this epidemiological burden is masked by under-reporting, which impairs planning of public policies for timely and effective intervention. An alternative system to improve the detection rate has been the active search for cases, especially in high risk groups, including homeless and incarcerated people. However, these activities require local resources, resulting in databases with different detection rates depending on the socio-economic characteristics and the management capacity of the municipalities. It is therefore crucial to estimate and quantify the uncertainty of the detection rates on a finer scale, to allow better informed decisions about the distribution of resources.

In this article we investigate a general framework for correcting under-reporting, suitable to a wide range of spatio-temporal count data, and apply it to counts of TB cases in Brazil. All counts can be potentially assumed under-reported (unlike other approaches) so that the severity of under-reporting is estimated and potentially informed by available covariates that relate to the under-reporting mechanism. The model is implemented in the Bayesian framework which allows great flexibility and leads to complete predictive distributions for the true counts, therefore quantifying the uncertainty in correcting the under-reporting. 

The article is structured as follows: Section \ref{sec:background} discusses approaches to modelling under-reporting, including the hierarchical framework we will ultimately use, as well as how we seek to resolve the incompleteness of the information provided by the data. Section \ref{sec:application} presents the application to Brazilian TB data, as well as some simulation experiments designed to investigate the sensitivity of the model's ability to quantify uncertainty. Further simulation experiments can be found in the Appendix, which address issues such as the sensitivity of the model to the strength of under-reporting covariates. Finally, Section \ref{sec:discussion} presents a critical evaluation of our approach, particularly compared to existing methods.

\section{Background}\label{sec:background}
Let $y_{i,t,s}$ be the number of events (e.g. TB cases) occurring in units of space $s\in S$, time $t\in T$ and any other grouping structures $i$ that the counts might be aggregated into. If $y_{i,t,s}$ is believed to have been perfectly observed, the counts are conventionally modelled by an appropriate conditional distribution $p(y_{i,t,s}\mid \bm{\theta})$, usually either Poisson or Negative Binomial. Here $\bm{\theta}$ represents random effects allowing for various dependency and grouping structures (e.g. space and time), as well as parameters associated with relevant covariates. Inference is then based on the conditional likelihood function (assuming independence in the $y_{i,t,s}$ given $\bm{\theta}$):
\begin{equation}\label{likelihood}
p(\bm{y}\mid \bm{\theta})=\prod_{i,t,s}p(y_{i,t,s} \mid \bm{\theta}).
\end{equation}

Under-reporting is conceptually a form of unintentional missing data \citep[ch.~8]{Gelman2013} where, in some or potentially all cases, we have not observed the actual number of events $y_{i,t,s}$. Instead, we have observed under-reported counts $z_{i,t,s}$, which represent lower bounds of $y_{i,t,s}$. This implies that using \eqref{likelihood} for all observed counts, under-reported or otherwise, will lead to biased inference. Rather, we should acknowledge the uncertainty caused by the missing $y_{i,t,s}$, whilst incorporating the partial information provided by the recorded counts $z_{i,t,s}$. More generally, the data collection mechanism should be included in the analysis and this is especially true for missing data problems. A conceptual framework is for this \citep[ch.~8]{Gelman2013} is one where both the completely observed (true) data and the mechanism determining which of them are missing are given probability models. Relating this more specifically to under-reporting, an indicator random variable $I_{i,t,s}$ is introduced, to index the data into fully observed or under-reported. In what follows, we review approaches to under-reporting that can be broadly classified into ones that treat $I_{i,t,s}$ as known, and ones that treat it as latent and therefore attempt to model it.

\subsection{Censored Likelihood}\label{sec:censored}
A common approach to correcting under-reporting is to base inference on the censored likelihood. This is the product of the evaluation of \eqref{likelihood} for the fully observed (uncensored) counts $y_{i,t,s}$ and the joint probability of the missing $y_{i,t,s}$ exceeding or equalling the recorded (censored) counts $z_{i,t,s}$:
\begin{equation}\label{Clikelihood}
p(\bm{y}\mid \bm{z}, \bm{\theta})=\prod_{I_{i,t,s}=1}p(y_{i,t,s}\mid \bm{\theta}) \prod_{I_{i,t,s}=0} p(y_{i,t,s} \geq z_{i,t,s} \mid \bm{\theta}).
\end{equation}
In this framework, the indicator $I_{i,t,s}$ for which data are under-reported is binary (where $I_{i,t,s}=1$ when $z_{i,t,s}$=$y_{i,t,s}$). The strength of this approach is that all of the observed counts contribute to the inference and, by accounting for the under-reporting in the model design, a more reliable inference on $\bm{\theta}$ is possible. However, information on which counts are under-reported is not always readily available, introducing the challenge of having to determine or estimate this classification. 

The approach in \cite{BAILEY2005335} accounts for under-reporting in counts of leprosy cases in the Brazilian region of Olinda, to arrive at a more accurate estimate of leprosy prevalence. They utilise prior knowledge on the relationship between leprosy occurrence rate and a measure of social deprivation to decide the values of $I_{i,t,s}$ a priori: A fixed value of social deprivation is chosen as a threshold, above which observations are deemed to be under-reported. However, the choice of this threshold is subjective and not always obvious. The approach can in principle be extended to include estimation of the threshold, however in many cases the threshold model may be a poor description of the under-reporting mechanism which could, for example, be related to more than one covariate. 

\cite{Oliveira2017} presents an alternative to this approach, which treats the binary under-reporting indicator $I_{i,t,s}$ as unobserved and therefore random. The classification of the data is characterised by $I_{i,t,s}\sim \mbox{Bernoulli}(\pi_{i,t,s})$, such that $\pi_{i,t,s}$ is the probability of any data point suffering from under-reporting, which is potentially informed by covariates. Although a more general approach in the sense of modelling the under-reporting classification, like any other censored likelihood method it lacks a way of quantifying the severity of under-reporting. This makes it unsuitable for our TB application, where we would like to learn about the under-reporting rate on a micro-regional level. Moreover, the predictive inference for the unobserved $y_{i,t,s}$ is limited, amounting to:
\begin{equation}
p(y_{i,t,s}\mid z_{i,t,s},\bm{\theta})=p(y_{i,t,s}\mid y_{i,t,s}\geq z_{i,t,s},\bm{\theta}).
\end{equation}
This is because the recorded counts $z_{i,t,s}$ are treated as constants, as opposed to random quantities arising jointly from the $y_{i,t,s}$ process and the under-reporting process. Therefore the severity of under-reporting does not contribute to the predictive inference. 

\subsection{Hierarchical count framework}\label{sec:hierarchical}
A potentially more flexible approach is to consider the under-reporting indicator variable $I_{i,t,s}$ as continuous in the range $[0,1]$, to be interpreted as the proportion of true counts that have been reported. This way, the severity of under-reporting is quantified and estimated when $I_{i,t,s}$ is assumed unknown. One way of achieving this is a hierarchical framework consisting of a Binomial model for the recorded counts $z_{i,t,s}$ and a latent Poisson model for the true counts $y_{i,t,s}$. This approach, often called the Poisson-Logistic \citep{pogit1993} or Pogit model, has been used across a variety of fields including economics (\cite{Winkelmann2008}, \cite{Winkelmann1996}), criminology \citep{MORENO}, natural hazards \citep{Stoner18} and epidemiology (\cite{GREER}, \cite{Dvorzak2016}, \cite{Shaweno2017}). The observed count $z_{i,t,s}$ is assumed a Binomial realisation out of an unobserved total (true) count $y_{i,t,s}$. The basic form of the model (extended in Section \ref{sec:application} to include spatial random effects) is given by:
\begin{eqnarray}
z_{i,t,s}\mid y_{i,t,s} &\sim& \text{Binomial}(\pi_{i,t,s},y_{i,t,s}) \label{z} \\  
\log\left(\frac{\pi_{i,t,s}}{1-\pi_{i,t,s}}\right) &=& \beta_0 + \sum_{j=1}^J\beta_jw^{(j)}_{i,t,s}\label{pi} \\
y_{i,t,s} &\sim& \text{Poisson}(\lambda_{i,t,s}) \label{y} \\
\log\left(\lambda_{i,t,s}\right) &=& \alpha_0 + \sum_{k=1}^K\alpha_kx^{(k)}_{i,t,s}\label{lambda}
\end{eqnarray}

All the data can be assumed to be (potentially) under-reported by treating $y_{i,t,s}$ as a latent Poisson variable in a hierarchical Binomial model for $z_{i,t,s}$. Assuming that all individual occurrences have equal chance of being independently reported, $\pi_{i,t,s}$ can be interpreted as the probability that each occurrence is reported, and is effectively the aforementioned indicator variable $I_{i,t,s}$. Relevant under-reporting covariates $\bm{W}=\{w^{(j)}_{i,t,s}\}$ (e.g. related to TB detection), enter the model through the linear predictor in the logistic transformation of $\pi_{i,t,s}$. This allows inference on the severity of under-reporting and what it relates to.
 
The true counts $y_{i,t,s}$ are modelled as a latent Poisson variable with mean $\lambda_{i,t,s}$, characterised (at the log-scale) as a linear combination of covariates $\bm{X}=\{x^{(k)}\}$ associated with the process giving rise to the counts. These are the covariates we would like to capture the effect of, or are known to influence $y_{i,t,s}$, including offsets such as population counts. In modelling TB incidence these include social deprivation indicators at a particular location. It is assumed that $\bm{W}$ and $\bm{X}$ are comprised of different variables so that the $w^{(k)}_{i,t,s}$ are unrelated to the process generating the counts.

Vectors $\bm{\alpha}=(\alpha_0,\ldots,\alpha_K)$ and  $\bm{\beta}=(\beta_0,\ldots,\beta_J)$ are parameters to be estimated. Using mean-centred covariates (column means of $\bm{X}$ and $\bm{W}$ are zero) implies that $\alpha_0$ and $\beta_0$ are respectively interpreted as the mean of $y_{i,t,s}$ on the log scale, and the mean reporting rate on the logistic scale, when the covariates are at their means. The framework allows the inclusion of random effects in both \eqref{pi} and \eqref{lambda}. Random effects allow for overdispersion in count models \citep[ch.~12]{Agresti2002}, and their inclusion here may be desirable to introduce extra variation and thus flexibility in the model for the true counts, including capturing effects from unobserved covariates. Alternatively, $y_{i,t,s}$ can be $\mbox{NegBin}(\lambda_{i,t,s},\theta)$: a Negative Binomial with mean $\lambda_{i,t,s}$ and dispersion parameter $\theta$ \citep{Winkelmann1998}. Moreover, some of the coefficients $\alpha_k$ could be assumed random to further increase model flexibility.

Considering the true counts as a latent variable aids in mitigating bias in estimating $\bm{\alpha}$ from under-reported data. The model is straightforward to implement in the conditional form \eqref{z}-\eqref{lambda}, by sampling $y_{i,t,s}$ using Markov Chain Monte Carlo (MCMC). However, doing so will likely result in slow-mixing MCMC chains that must be run for a large number of iterations to achieve a desired effective sample size. Conveniently the following two results are achieved by integration and use of Bayes' rule:
\begin{eqnarray}
z_{i,t,s} \sim \text{Poisson}(\pi_{i,t,s}\lambda_{i,t,s})\label{integrated}\\ 
y_{i,t,s} - z_{i,t,s}  \sim \text{Poisson}((1-\pi_{i,t,s})\lambda_{i,t,s}) \label{difference}
\end{eqnarray}
If $y_{i,t,s}\sim \mbox{NegBin}(\lambda_{i,t,s},\theta)$, then $z_{i,t,s}\sim\mbox{NegBin}(\pi_{i,t,s}\lambda_{i,t,s},\theta)$. The consequence of this is that the model in \eqref{integrated} is much more efficient in terms of effective sample size per second, while samples of $\bm{y}$ can be generated using Monte Carlo simulation of \eqref{difference}. This also means that a complete predictive inference on the true counts $y_{i,t,s}$ is possible, deriving information jointly from the mean rate of $y_{i,t,s}$, the reporting probability $\pi_{i,t,s}$ and the recorded counts $z_{i,t,s}$.

However, equation \eqref{integrated} suggests that the same observed counts $z_{i,t,s}$ could arise from either a high $\lambda_{i,t,s}$ value combined with a low $\pi_{i,t,s}$, or vice versa, so that the likelihood function of $z_{i,t,s}$ is constant over the level curves of $\pi_{i,t,s}\lambda_{i,t,s}$. This means that, in the absence of any completely reported observations, there is a lack of identifiability between the two intercepts $\alpha_0$ and $\beta_0$. Additionally, as illustrated in Appendix \ref{app:class}, the framework cannot automatically identify whether a given covariate is associated with the under-reporting or the count generating process. This means that care must be taken when deciding which part of the model a covariate belongs in. Non-identifiability for models where the mean is a product of an exponential and logistic term is discussed in greater detail by \cite{identifiability2012}, with discussion more specific to under-reporting in \cite{Papadopoulos2008}. 

To conduct meaningful inference on the true counts $y_{i,t,s}$, the partial information in the data must be supplemented with extra information to differentiate between under-reporting and true incidence rate. One potential source of information is to utilise a set of completely reported observations alongside the potentially under-reported observations, an approach used by \cite{Dvorzak2016} and \cite{STAMEY}. For these counts, the reporting probability $\pi_{i,t,s}$ (and hence the indicator variable $I_{i,t,s}$) is known a priori to equal 1. In practice, this can be implemented by replacing \eqref{pi} with:
\begin{eqnarray}
\pi_{i,t,s}= c_{i,t,s} + (1-c_{i,t,s})\exp{\bigg\{\frac{\eta_{i,t,s}}{1+\eta_{i,t,s}}\bigg\}}
\end{eqnarray}
Here $c_{i,t,s}$ is an indicator variable, where $c_{i,t,s}=1$ when $z_{i,t,s}$ is completely reported ($\pi_{i,t,s}=1$) and 0 otherwise ($\pi_{i,t,s}$ is unknown), and $\eta_{i,t,s}$ is the right hand side of \eqref{pi}. For some applications, however, such as historical counts of natural hazards \citep{Stoner18}, it is often impractical and even impossible to obtain completely observed data. For the application to Brazilian TB data in Section \ref{sec:application}, complete counts of cases are not available on a micro-regional level. An alternative source of information \citep{MORENO} is to employ informative prior distributions to differentiate between $\pi_{i,t,s}$ and $\lambda_{i,t,s}$, which is the approach we adopt in modelling TB. In Appendix \ref{app:information}, we examine the effects of either source of information on prediction uncertainty using simulation experiments.

Recently, \cite{Shaweno2017} applied a version of this framework to TB data in Ethiopia, without any data identified as completely observed. However, vague uniform priors are used for regression coefficients, including the intercepts $\alpha_0$ and $\beta_0$. Because of this ambiguity as to whether in practice it is necessary to use an informative prior distribution, we also conduct a thorough investigation of the sensitivity of the framework to the choice of prior distributions using simulated data, in Section \ref{sec:sims}.

In summary, the strengths of the hierarchical count framework over the more traditional censored likelihood approach are that it allows both for varying severity of under-reporting across data points and for a more complete predictive inference on the true counts.
\section{Model Application}\label{sec:application}
\begin{figure}[h!]
\floatbox[{\capbeside\thisfloatsetup{capbesideposition={right,center},capbesidewidth=2.5in}}]{figure}[\FBwidth]
{\caption{Total new TB cases for each mainland micro region of Brazil, over the years 2012-2014, per 100,000 inhabitants.} \label{fig:data}}
{\includegraphics[scale=0.6]{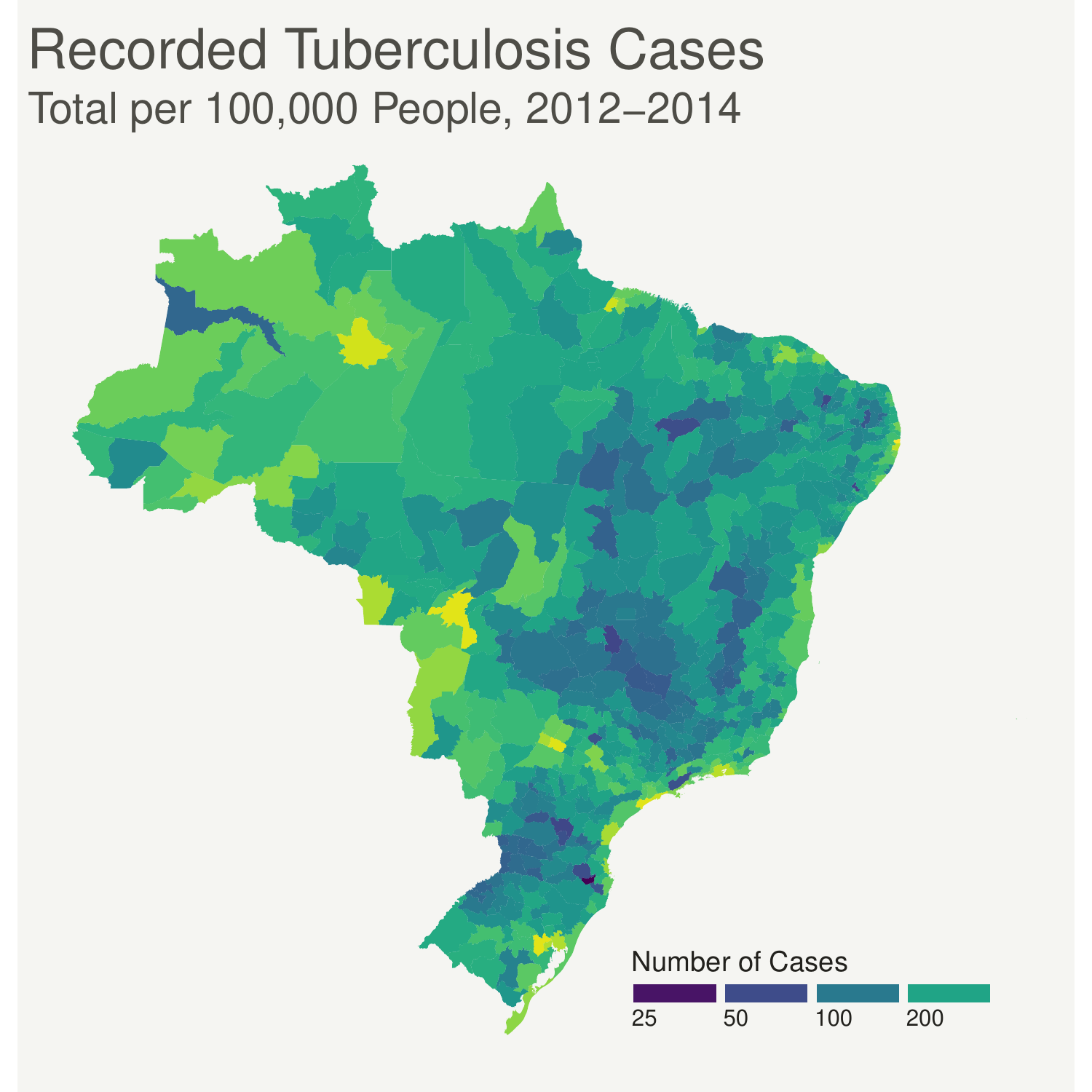}}
\end{figure}

Let $y_{t,s}$ and $z_{t,s}$ denote respectively the true and recorded counts of TB cases in micro region $s\in \{1,\ldots,557\}$ (spanning all of Brazil), and year $t\in \{2012,2013,2014\}$. Figure \ref{fig:data} illustrates the recorded TB incidence rate. A spatial structure is apparent, with generally higher TB rates in the north-west than in the south-east. Some of this variability may be attributed to spatial covariates affecting TB incidence. In particular, high risk populations include poorly integrated groups due to poverty related issues, such as homelessness and incarceration. To allow for this, various social deprivation indicators for each micro-region were considered as covariates. These were: $x^{(1)}_s=$ unemployment (the proportion of economically active adults without employment); $x^{(2)}_s=$ urbanisation (the proportion of people living in an urban setting); $x^{(3)}_s=$ density (the mean number of people living per room in a dwelling); and $x^{(4)}_s=$ indigenous (the proportion of the population made up by indigenous groups).
\newpage
Furthermore, the covariate $u_s=$ treatment timeliness (the proportion of TB cases for which treatment begins within one day) was considered in the characterisation of the under-reporting mechanism. Having already controlled for social deprivation through $x^{(j)}_s$, $u_s$ acts as a proxy for how well a local TB surveillance programme is resourced. The model is specified (conditionally on random effects) as follows:
\begin{eqnarray}
z_{t,s}\mid y_{t,s},\gamma_{t,s} &\sim& \text{Binomial}\left(\pi_{s},y_{t,s}\right) \\
\log\left(\frac{\pi_{s}}{1-\pi_{s}}\right)&=&\beta_0 + g(u_s) + \gamma_{t,s} \label{pi3} \\
y_{t,s}|\phi_s,\theta_s &\sim& \text{Poisson}(\lambda_{t,s}) \\
\log\left(\lambda_{t,s}\right) &=& \log(P_{t,s}) + a_0 + f_1(x^{(1)}_s) +  f_2(x^{(2)}_s) \nonumber \\
&+& f_3(x^{(3)}_s) + f_4(x^{(4)}_s) + \phi_s + \theta_s \label{lambda3}
\end{eqnarray}
Functions $g(\cdot)$, $f_1(\cdot),\ldots,f_4(\cdot)$ are orthogonal polynomials of degrees 3, 2, 2, 2 and 1, respectively. Compared to raw polynomials, these reduce multiple-collinearity between the monomial terms \citep{Kennedy}, and were set up using the ``poly'' function in R \citep{R}. The polynomials are defined such that $f(x)=0$ when $x=\bar{x}$, so that (at the logistic scale) $\beta_0$ is the mean reporting rate for a region with mean treatment timeliness. The term $\log(P_{t,s})$, where $P_{t,s}$ is population, is an offset to allow for varying population and ensure the covariates act on the incidence rate.

Additive effects from a spatially unstructured random effect $\theta_s$ and a spatially structured one, $\phi_s$ are assumed to capture any residual spatial variation in the incidence of TB. An Intrinsic Gaussian Conditional Autoregressive (ICAR) model \citep{Besag1991} was assumed for $\phi_s$, with variance parameter $\nu^2$, to capture dependence between neighbouring micro-regions. Here, a neighbour of $s$ was defined as any $s'\neq s$ sharing a geographical boundary with $s$. The $\mbox{N}(0,\sigma^2)$ effect $\theta_s$ was included to afford extra spatial residual variability. An additional unstructured $\mbox{N}(0,\epsilon^2)$ effect $\gamma_{t,s}$ was included in the model for the reporting rate \eqref{pi3}, to allow for the effect of potential unobserved covariates on the detection rate of TB, as well as the case that $u_s$ may only be a proxy for the appropriate (true) under-reporting covariate.

The prior distribution for $\alpha_0$ was assumed $\mbox{N}(-8,1)$, chosen by using prior predictive checking to reflect our belief that very high values (such as over 1 million) for the total number of cases are unlikely. The priors for $\alpha_j$ $(j=1,...,7)$ and $\beta_k$ $(k=1,2,3)$ were specified as $\mbox{N}(0,10^2)$, which were chosen to be relatively non-informative. Finally, the priors for variance parameters $\sigma$, $\nu$ and $\epsilon$ were specified as zero-truncated $\mbox{N}(0,1)$, to reflect the belief that low variance values are more likely than higher ones, but that these effects are likely to capture at least some of the variance. As discussed in Section \ref{sec:hierarchical}, in the absence of any completely reported TB counts, we must specify an informative prior distribution for $\beta_0$ to supplement the partial information in the data. As an aid in doing so, we investigate the sensitivity of the model to this prior through simulation experiments presented in the following subsection. 

All models were implemented using NIMBLE \citep{nimble}, a facility for flexible implementations of MCMC models in conjunction with R \citep{R}. Specifically, we made use of the Automated Factor Slice Sampler (AFSS) which can be an efficient way of sampling vectors of highly correlated parameters \citep{AFSS}, such as $\alpha_0$ and $\beta_0$. The associated code and data are provided as supplementary material. 

\subsection{Simulation experiments}\label{sec:sims}
\begin{figure}[h!]
\centering
\caption{Scatter plots of simulated data, showing the process covariate $x_s$ against the true counts $y_s$ (left), the under-reporting covariate $w_s$ against $y_s$ (centre) and $w_s$ against the recorded counts $z_s$ (right).}
\label{fig:sim_values}
\includegraphics[width=\linewidth]{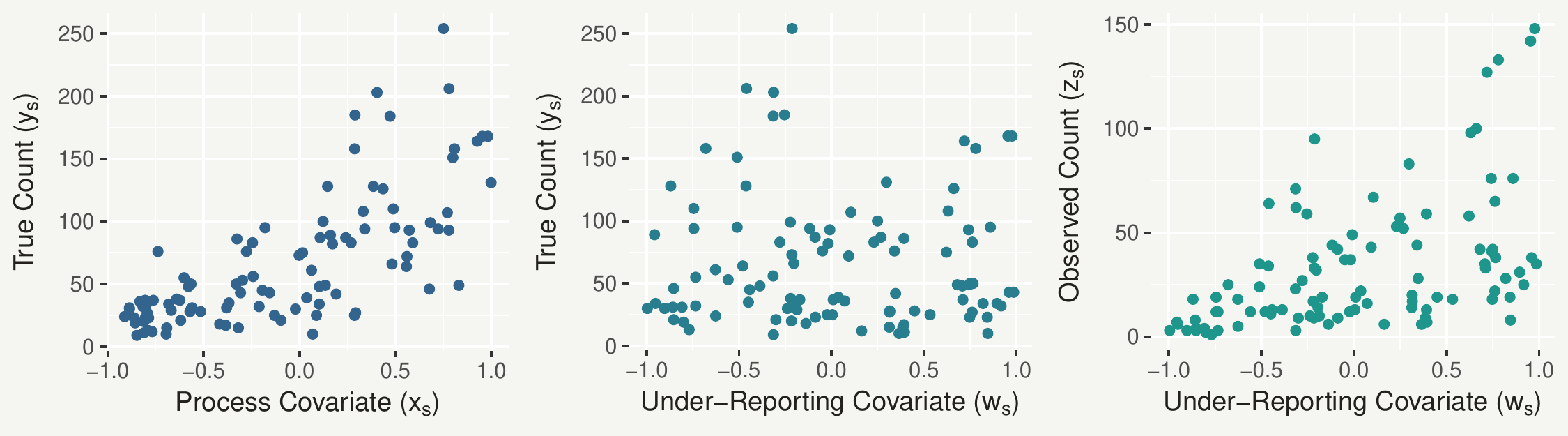}
\end{figure}

For the simulation study, we consider counts which vary in space in the following way: 
\begin{eqnarray}
z_{s}|y_{s} &\sim& \text{Binomial}(\pi_{s},y_{s}) \\
\log\left(\frac{\pi_{s}}{1-\pi_{s}}\right) &=& \beta_0 + \beta_1 w_{s}  \label{truepi} \\ 
y_{s}|\phi_{s} &\sim& \text{Poisson}(\lambda_{s})\\
\log\left(\lambda_{s}\right) &=& \alpha_0 + \alpha_1 x_{s} + \phi_{s} \label{truelambda} 
\end{eqnarray}
with $\beta_0=0$, $\beta_1=2$, $\alpha_0=4$, $\alpha_1=1$ and $\nu=0.5$. A total of $s=1,\ldots,100$ data points were simulated with both covariates $x_{s}$ and $w_{s}$ being sampled from a $\mbox{Unif}(-1,1)$ distribution. The $\mbox{ICAR}(\nu^2)$ spatial effect $\phi_s$ was simulated over a regular 10x10 lattice. Figure \ref{fig:sim_values} shows the simulated data. Note there are clear positive relationships between $x_{s}$ and $y_{s}$, and between $w_{s}$ and $z_{s}$, while there is no clear relationship between $w_s$ and $y_s$. One goal for this simulation is to investigate the sensitivity of the model to the specification of the Gaussian prior distribution for $\beta_0$. This was achieved by repeatedly applying the model whilst varying the mean and standard deviation for this prior. The prior for $\alpha_0$ was $\mbox{N}(0,10^2)$, with all other priors the same as in the TB model.

To make the experiment more realistic, we mimic the case where the true under-reporting covariate $w_{s}$ is not available, and instead we only have access to (proxy) covariates $v_{s,2},...,v_{s,6}$. These are simulated such that they have decreasing correlation with $w_{s}$. As the variation in $\pi_s$ is no longer fully captured by $v_{s,2},...,v_{s,6}$, we include a random quantity $\gamma_s\sim\mbox{N}(0,\epsilon^2)$ in \eqref{truepi}.

\begin{figure}[h!]
\floatbox[{\capbeside\thisfloatsetup{capbesideposition={right,center},capbesidewidth=2.5in}}]{figure}[\FBwidth]
{\caption{Coverage of the 95\% PIs for $y_s$, when the under-reporting covariate $v_{s,3}$, which has a theoretical correlation of 0.6 with the true covariate $w_s$, is used.} \label{fig:coverage}}
{\includegraphics[scale=0.8]{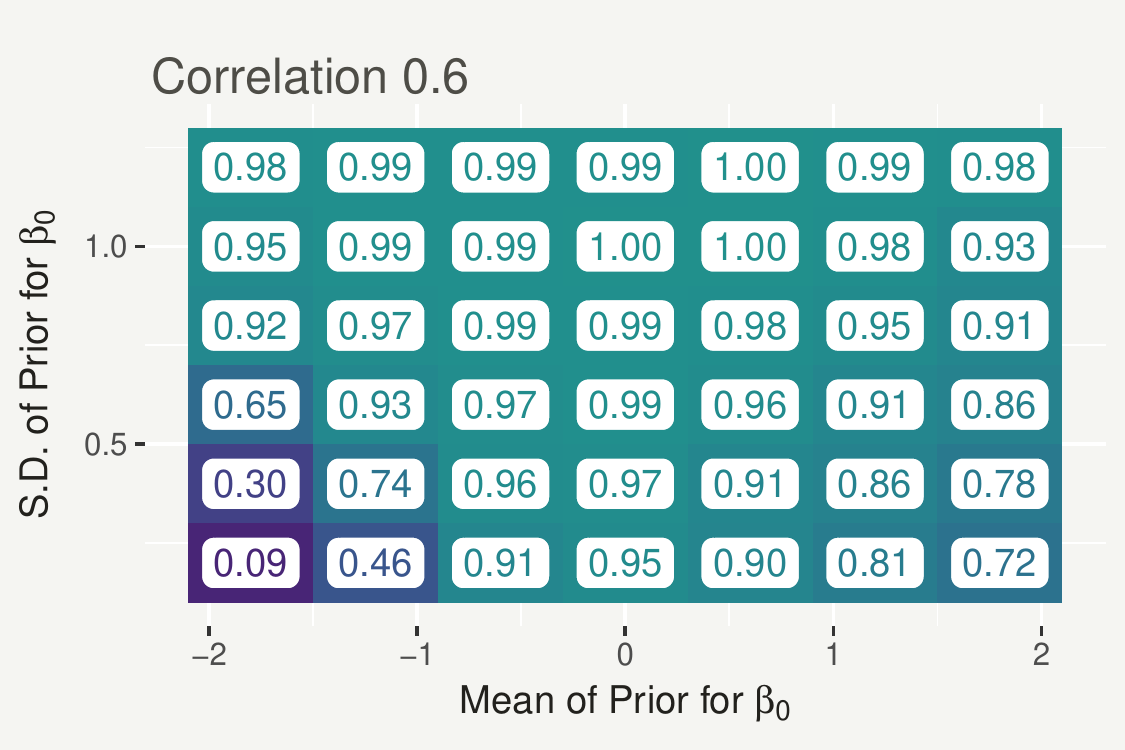}}
\end{figure}

An important aspect of model performance to consider is the proportion of true counts that lie in their corresponding 95\% posterior prediction intervals (PIs), known as the coverage. In the context of non-identifiability, we would expect the coverage to remain high as long as the true value of $\beta_0$ is not extreme with respect to its prior. Figure \ref{fig:coverage} shows the coverage when the covariate $v_{s,3}$ (correlation 0.6 with $w_s$) is used (which incidentally has a similar correlation value with the recorded counts as treatmeant timeliness in the TB data). The plot suggests that the model is able to quantify uncertainty well, as long as a strong prior distribution is not specified well away from the true value (lower corners). The inclusion of $\gamma_s$ implies that using a ``weaker'' under-reporting covariate should have little impact on coverage (the PIs of $y_s$ would simply widen). Indeed, more detailed results in Appendix \ref{app:strength} show that mean coverage did not change systematically when weakening the covariate.

\begin{figure}[h!]
\centering
\caption{The top-left, top-central and lower-left plots show density estimates of prior (black) and posterior (coloured) samples for parameters $\alpha_0$, $\alpha_1$ and $\beta_0$, respectively, with vertical lines representing their true values. The top-right plot shows the mean predicted spatial effect ($\phi_s$) against the true values. The lower-central plot shows the predicted relationship (solid line) between the under-reporting covariate $v_{s,3}$ and the reporting probability $\pi_s$, with associated 95\% CrI. The lower-right plot shows the lower (blue) and upper (green) limits of the 95\% PIs for the true counts $y_s$.}
\label{fig:par_post}
\includegraphics[width=\linewidth]{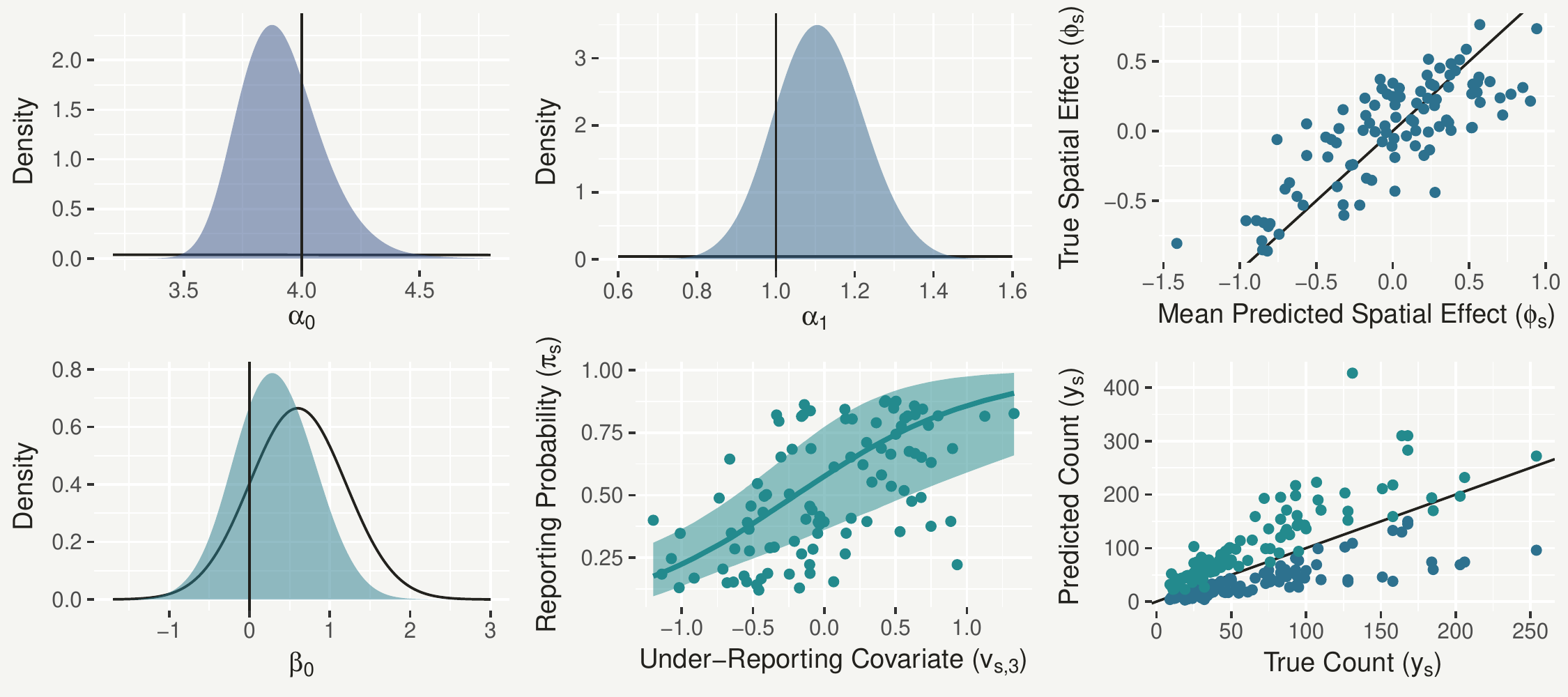}
\end{figure}

As an illustrative example of model performance, Figure \ref{fig:par_post} shows various results based on simulated data using $v_{s,3}$ as the under-reporting covariate, and a $\mbox{N}(0.6,0.6^2)$ prior for $\beta_0$. This represents the case where the prior distribution overestimates the reporting probability but not to an extreme extent. The top left and central plots show posterior densities for $\alpha_0$ and $\alpha_1$, indicating substantial learning of these parameters compared to the flat priors also shown. The top right plot compares the mean predicted spatial effects to their corresponding true values, suggesting these are captured well. The lower-left plot shows the posterior for $\beta_0$ has shifted in the direction of the true value. This illustrates that, at least in this idealised setting, the model is not entirely at the mercy of the accuracy of this prior, despite non-identifiability. The bottom central plot shows the mean predicted effect of the imperfect covariate $v_{s,3}$ on the reporting probability, with associated 95\% credible interval (CrI). The effect is quite uncertain, reflecting the relative weakness of the covariate. Finally, the lower right plot shows the lower (blue) and upper (green) limits of the 95\% PIs for $y_s$, suggesting that the model is able to systematically predict well the true unobserved counts.

This sensitivity analysis is by no means exhaustive, but it does appear to suggest that the model with no completely observed values is robust in terms of quantifying uncertainty, as long as the practitioner specifies a prior for $\beta_0$ that is informative but not too strong. With this in mind, we return to the task of specifying this prior distribution for the TB model. The information available are WHO inventory study-derived estimates \citep{TBInventory} of the overall TB detection rate in Brazil for 2012-2014. The 2017 point estimates for these years, with associated 95\% confidence intervals were 91\% (78\%,100\%), 84\% (73\%,99\%) and 87\% (75\%,100\%) \citep{TBReport}. Normal distributions were used to approximate each rate at the logistic level. We inferred mean and standard deviation parameter values by attempting to match the quoted point estimates and confidence intervals. The mean of the three rates is most variable when they are positively correlated, so to account for this we simulated and sorted into ascending order samples from each approximate distribution, before computing the mean of each sample of three rates. This resulted in a distribution which was approximately $\mbox{N}(2,0.4^2)$. Figure \ref{fig:coverage} suggests that the mean of this prior can only be slightly wrong (less than 0.5 away) before coverage begins to drop below ideal levels (95\%). For this reason, and because the incorporation of the WHO uncertainty is only approximate, we opt for a more conservative standard deviation of 0.6, which allows the mean to deviate more from the truth before PIs become less trustworthy.

\subsection{Model checking}\label{sec:check}
As well as inspecting trace plots of MCMC samples, convergence was assessed by computing the potential scale reduction factor (PSRF) for each parameter \citep{convergence}, which compares the between-chain and within-chain variances. If the chains have not converged, the between-chain variance should exceed the within-chain variance and the PSRF will be substantially greater than 1. Using different initial values and random number seeds for each chain gives the best assurance that the chains have converged to the whole posterior, rather than a local mode. Four chains were used, each ran for a total of 800K iterations. After discarding 400K iterations as burn-in, the PSRF was computed as less than 1.05 for all regression coefficients and variance parameters. These were deemed sufficiently close to 1 to indicate convergence.

A natural way of assessing whether the model fits the data well is to conduct posterior predictive model checking \citep[ch.~6]{Gelman2013}. More specifically, one can look at the discrepancy between the data $\bm{z}$ and posterior predictive replicates of this data from the fitted model. Define the posterior predictive distribution for a replicate $\tilde{z}_{t,s}$, of observed number of TB cases $z_{t,s}$, as $p(\tilde{z}_{t,s}\mid \bm{z})$. The question is then whether the actual observation $z_{t,s}$ is an extreme value with respect to $p(\tilde{z}_{t,s}\mid \bm{z})$ and if so, this indicates poor model performance.
\begin{figure}[h!]
\floatbox[{\capbeside\thisfloatsetup{capbesideposition={right,center},capbesidewidth=6cm}}]{figure}[\FBwidth]
{\caption{Scatter plot of differences between the lower (blue) and upper (green) limits of the 95\% PIs of $\tilde{z}_{t,s}$ and the observed values $z_{t,s}$.}\label{fig:test}}
{\includegraphics[scale=0.8]{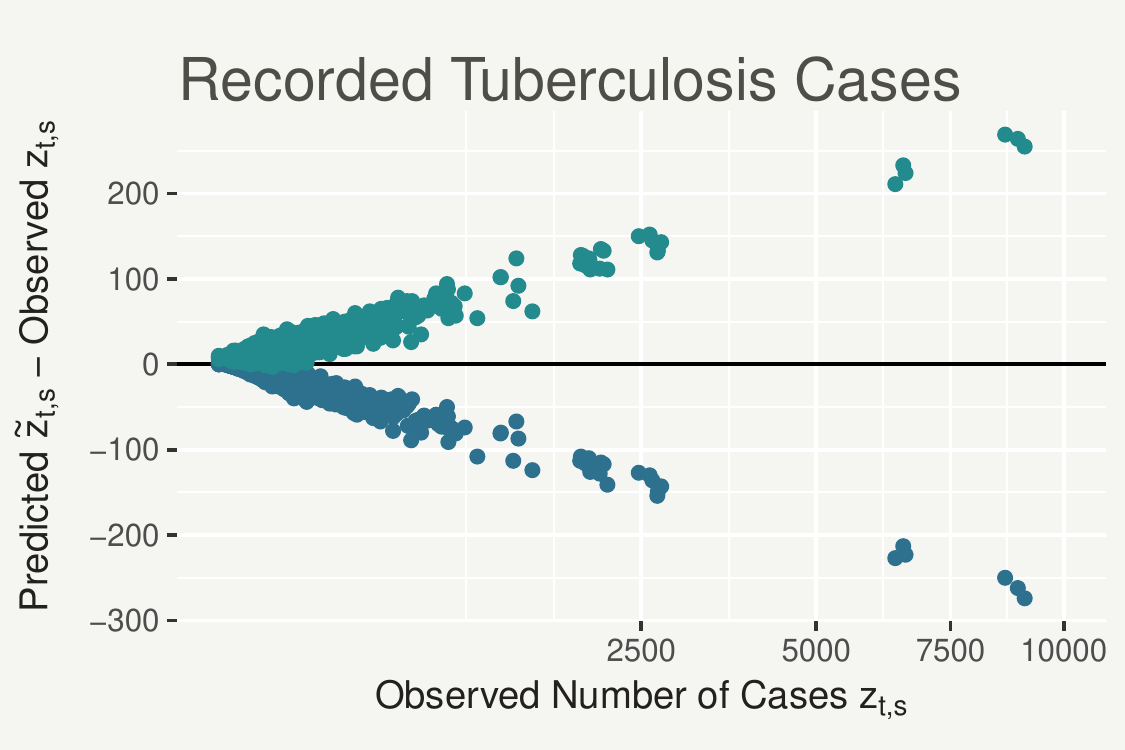}}
\end{figure}

Figure \ref{fig:test} shows a scatter plot of the difference between the lower (blue) and upper (green) limits of the 95\% posterior PIs of $\tilde{z}_{t,s}$ and the corresponding observed values $z_{t,s}$. The PIs are symmetrically centred on the observed values, suggesting that the model has no systematic issue (under or over-prediction) with fitting observed values. The coverage of the 95\% PIs was approximately 99.6\%.

Furthermore, we can assess whether summary statistics of the original data are captured well by the model through the replicates. Given this is count data, we want to ensure that both the sample mean and variance are captured well. As the prior distributions used for regression coefficients were quite broad, it is important to also assess whether substantial learning has occurred, with respect to both the predictive error of the observed counts $z_{t,s}$ and the distributions of these statistics. Otherwise, it is possible that the data are well captured in the posterior predictions because they were contained within the prior predictions.
\begin{figure}[h!]
\caption{Prior (top row) and posterior (bottom row) predictive distributions of the sample mean (left column), sample variance (central column) and the log-mean squared error from the recorded counts $z_{i,t,s}$ (right column), of the replicates $\tilde{z}_{t,s}$. Observed statistics are plotted as vertical lines.} \label{fig:check}
\includegraphics[width=\linewidth]{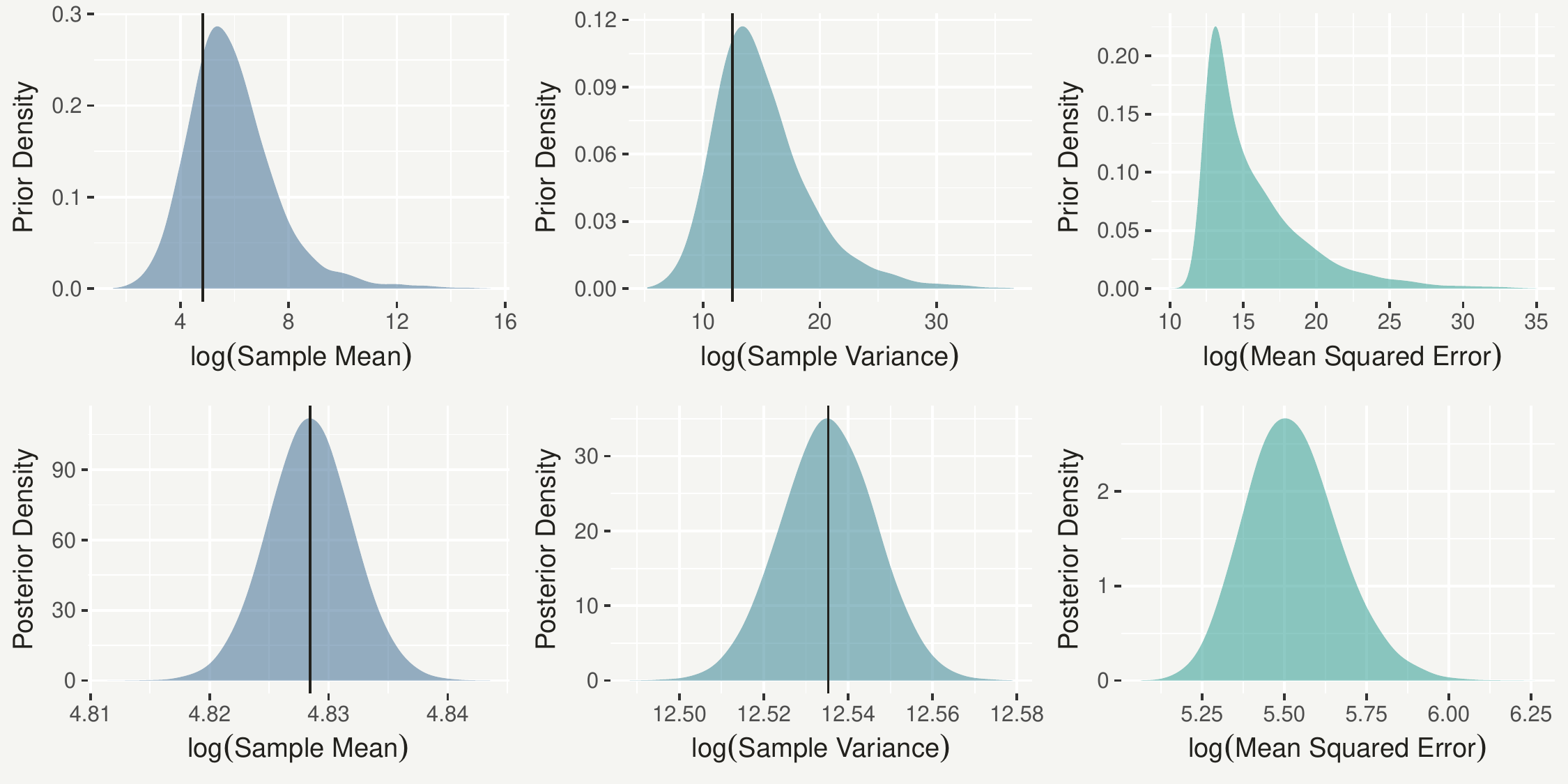}
\end{figure}

The left and central columns of Figure \ref{fig:check} show the prior (top) and posterior (bottom) predictive distributions of the sample mean and variance. The corresponding observed quantities are in the bulk suggesting that the prior and posterior models capture these well. The posterior predictive distributions are far more precise, indicating that the uncertainty in the parameters has been reduced significantly by the data. This is emphasised by the right column, which compares the posterior and prior predictive distributions of the mean squared difference between each $\tilde{z}_{t,s}$ and $z_{t,s}$. The mean squared error is several orders of magnitude smaller in the posterior model, implying far greater prediction accuracy.

\subsection{Results}\label{sec:results}
\begin{figure}[h!]
\includegraphics[width=\linewidth]{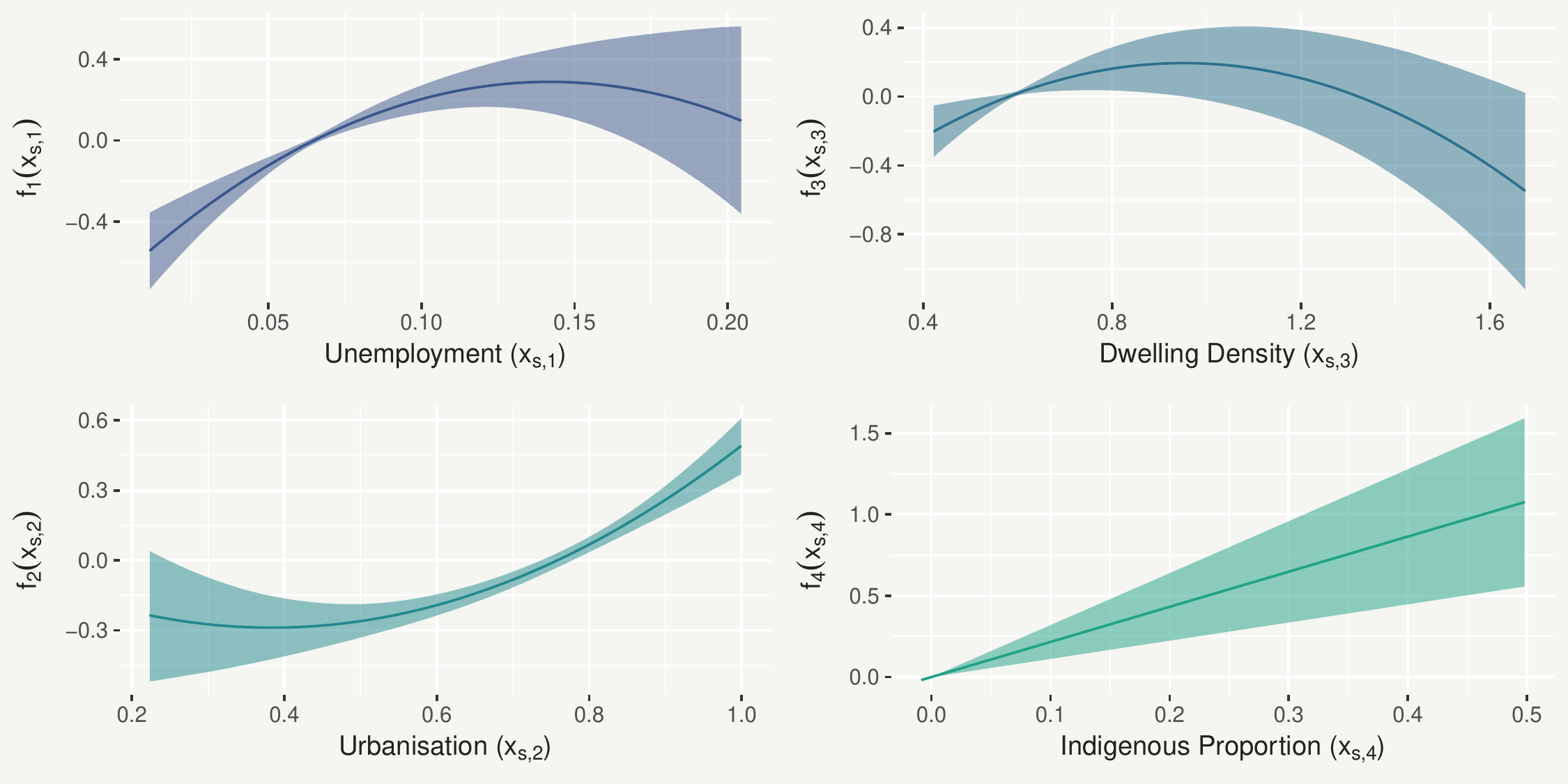}
\caption{Posterior mean predictions (solid lines) of the effects of unemployment, indigenous, density and urbanisation on the rate of TB incidence, with associated 95\% CrIs.} \label{fig:fplots}
\end{figure}

The effect of unemployment on $\lambda_{t,s}$ is shown in the upper-left panel of Figure \ref{fig:fplots}, indicating a strong (based on the width of the 95\% CrIs) positive relationship with TB incidence. This is likely because areas with high unemployment often also have high rates of homelessness and incarceration, two important risk factors for TB. The range of this effect is approximately 0.8 on the log scale, suggesting incidence rate is over twice as high in micro-regions with high unemployment ($>15\%$), compared to areas with low unemployment ($<5\%$). The lower-left panel shows that urbanised proportion is also strongly positively related to TB incidence. The range of this effect is also approximately 0.8, meaning that highly urbanised ($>90\%$) micro-regions are predicted to have over double the TB incidence of micro-regions with low urbanisation ($<40\%$). This could be due to the increased population density of highly urbanised areas, which may promote the spread of the disease. The effect of dwelling density is less pronounced: the polynomial increases monotonically for most of the range covered by the data ($x_s^{(3)}$ $<1$), before decreasing for higher values. This suggests that TB incidence is actually lower in micro-regions with the highest levels of dwelling density. Alternatively it may be that further under-reporting of TB is present in such areas, which is not being captured by this model. Data at these upper values are quite sparse, as reflected by widening of the 95\% CrIs. Finally, the lower-right panel of Figure \ref{fig:fplots} shows the effect of indigenous proportion. Recall that this relationship was constrained to be linear in \eqref{lambda3} and the 95\% CrI on the slope suggests the effect is strongly positive.

\begin{figure}[h!]
\floatbox[{\capbeside\thisfloatsetup{capbesideposition={right,center},capbesidewidth=2.5in}}]{figure}[\FBwidth]
{\caption{Combination of structured spatial effect $\phi_s$ and unstructured effect $\theta_s$.} \label{fig:spatial}}
{\includegraphics[scale=0.6]{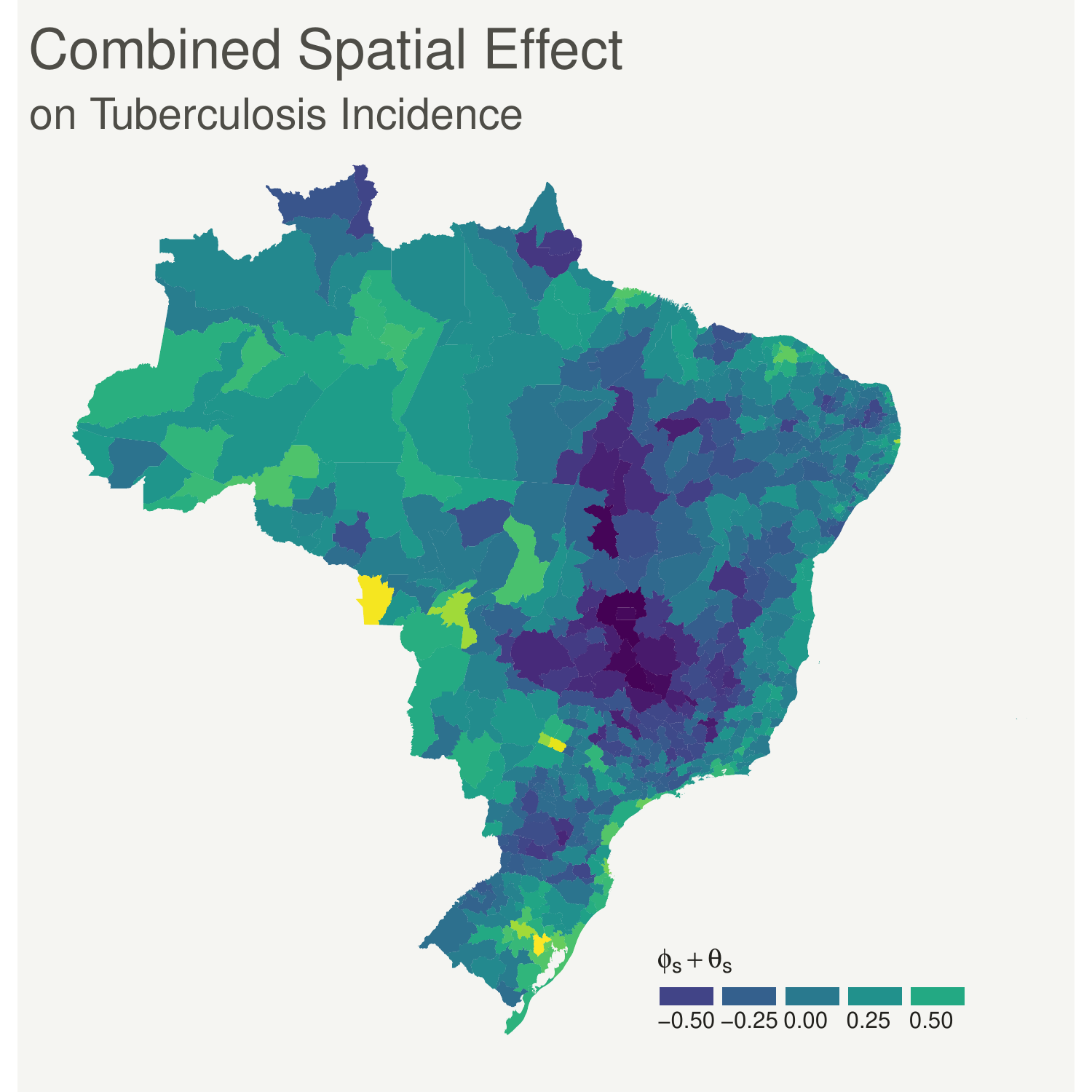}}
\end{figure}

\newpage
Figure \ref{fig:spatial} illustrates the predicted residual spatial variability in the TB incidence rate ($\phi_s+\theta_s$). There is substantial clustering of negative values in the centre of Brazil, surrounding the states of Goi\'{a}s and Tocantins, while there is clustering of positive values in the North West, including the Amazon rainforest. Interestingly, this seems to align well with estimates of the spatial distribution of human development index (HDI) (see for instance \cite{Atlas}), where high estimates of HDI coincide with low values from the spatial effect. This could indicate that there exist other effects of human development on TB incidence, such as healthcare infrastructure, which are not captured by the covariates. Several big cities, including Rio de Janeiro and S\~{a}o Paulo appear to buck this trend, with positive spatial effects despite relatively high HDI estimates, which could be due to the effect of features unique to big cities, such as high population density, which aren't included in the model. The effect of the spatially structured $\phi_s$ is visible by the clustering of similar colours and we found it dominated the unstructured effect $\theta_s$, explaining a predicted 94\% of their combined variation. The range of values of the combined effect is not dissimilar to the effects of any of the individual covariates, implying that the covariates are driving most of the variability in the true counts $y_{t,s}$.

\begin{figure}[h!]
\floatbox[{\capbeside\thisfloatsetup{capbesideposition={right,center},capbesidewidth=2.6in}}]{figure}[\FBwidth]
{\caption{Posterior mean predicted effect of treatment timeliness on the reporting probability of TB, with associated 95\% CrI.}\label{fig:timeliness}}
{\includegraphics[scale=0.8]{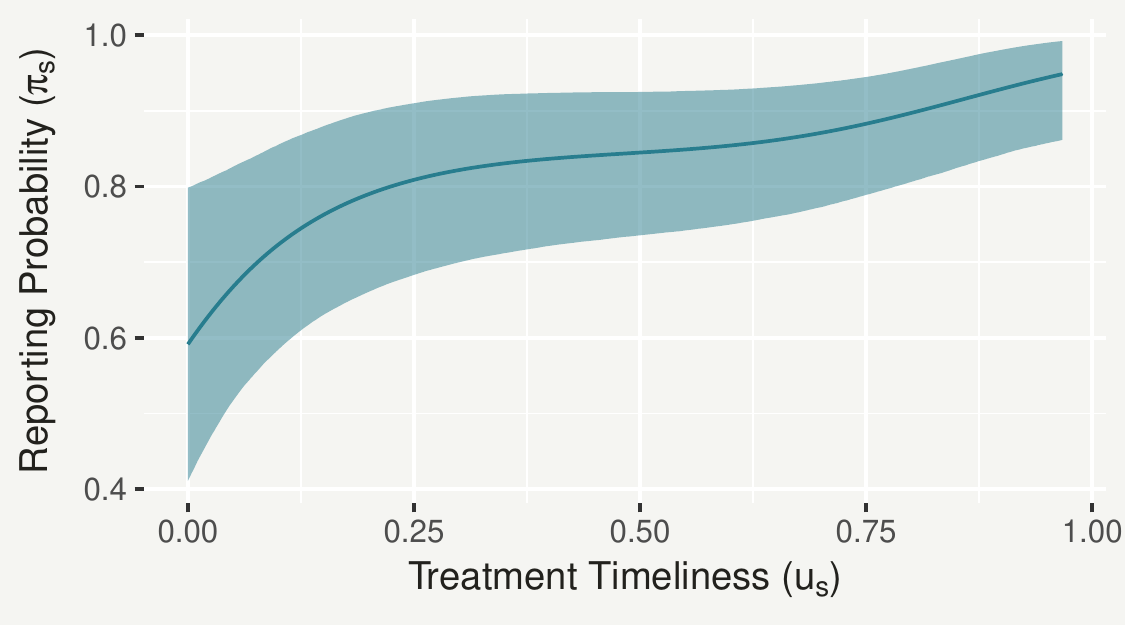}}
\end{figure}

\begin{figure}[h!]
\floatbox[{\capbeside\thisfloatsetup{capbesideposition={right,center},capbesidewidth=2.6in}}]{figure}[\FBwidth]
{\caption{Bar plot showing, for each year, the recorded total number of TB cases in Brazil, as well as the 5\%, 50\% and 95\% quantiles of the predicted true total number of TB cases.} \label{fig:totalbar}}
{\includegraphics[scale=0.8]{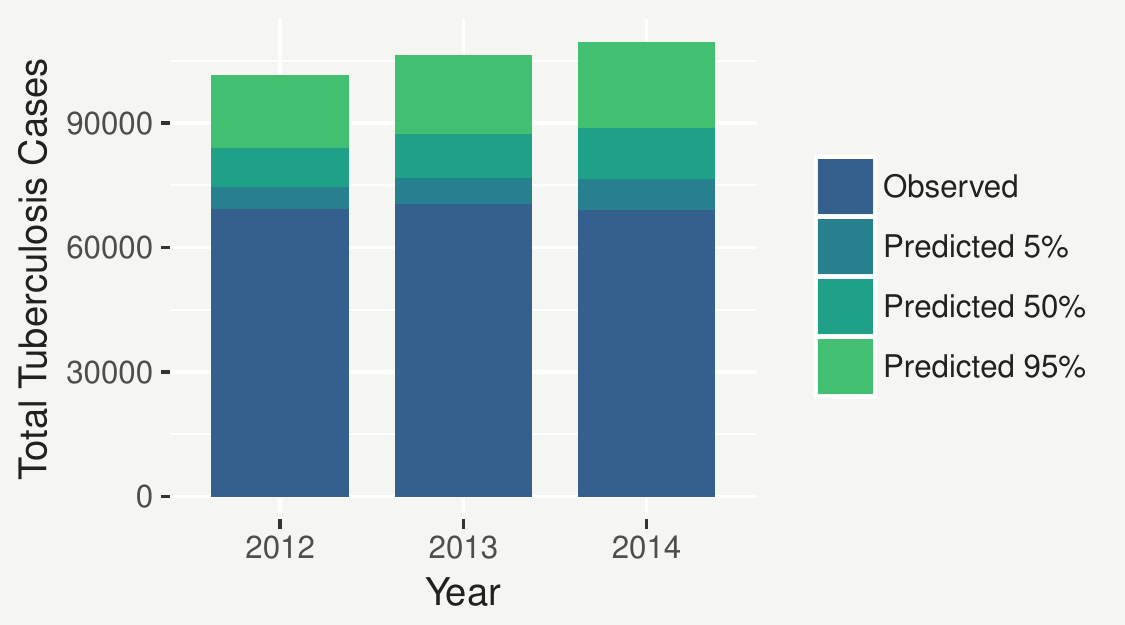}}
\end{figure}

Figure \ref{fig:timeliness} shows a clear, monotonically increasing (estimated) relationship between treatment timeliness and the probability of reporting $\pi_{t,s}$. The 95\% CrI does not incorporate a horizontal line, which would imply no relationship. Overall, micro-regions with very low timeliness ($<10\%$) have approximately two-thirds the reporting probability of ones with very high timeliness ($>90\%$), indicating a clear disparity in the performance of the surveillance programs.
\newpage
Finally, Figure \ref{fig:totalbar} shows, for each year, the total observed TB count, alongside the 5\%, 50\% and 95\% quantiles of the predicted true total number of unreported cases. The plot suggests that potentially tens of thousand of cases went unreported each year. Combined with the results seen in Figure \ref{fig:timeliness}, this presents a strong case for providing additional resources to the surveillance programs in those micro-regions with lower values of treatment timeliness. The R code and data needed to reproduce these results are provided as supplementary material.

\section{Discussion}\label{sec:discussion}
A flexible modelling framework for analysing potentially under-reported count data was presented. This approach can accommodate a situation where all the data are potentially under-reported, by using informative priors on model parameters which are easily interpretable. It also readily allows for random effects for both the disease incidence process and the under-reporting process, something which simulation experiments revealed alleviates the use of proxy covariates to determine under-reporting rates. It was applied to correcting under-reporting in TB incidence in Brazil using well-established MCMC software, incorporating a spatially structured model which highlights its flexibility. Simulation experiments were conducted to investigate prior sensitivity and to provide a guide for choosing a prior distribution for the mean reporting rate. 

Naturally, care should be taken. Indeed, it is likely that a different prior distribution for $\beta_0$ in the TB application might result in different inference on the under-reporting rate, and consequently the corrected counts. The simulation experiments indicated that if the specified prior information on the overall under-reporting rate turns out to be wildly different from the truth, then the corrected counts will also likely be inaccurate. Therefore particular attention should be paid to the elicitation of this prior information, such that the prior uncertainty is fully quantified and reflected in predictive inference. Further simulation experiments also highlighted the risk posed by incorrectly classifying covariates as either belonging in the under-reporting mechanism or the model of the true count. In many cases strong prior information about this classification may be available, so we suggest future research is directed at combining prior uncertainty with methods such as Bayesian model averaging. This could more rigorously quantify the uncertainty associated with this classification and its effect on the predictive inference for the corrected counts.

The subjective nature of the solution to completely under-reported data is not unique; in \cite{BAILEY2005335} for example, a different choice of threshold for the variable used to identify under-reported counts could have lead to different predictions. Only the usage of a validation study (e.g. \cite{STAMEY}) could be considered a less subjective approach depending on the quality, quantity and experimental design of collecting the validation data. In many cases however, the elicitation of an informative prior distribution for one parameter is simply a more feasible solution. In the application to TB, an existing estimate from the WHO of the overall reporting rate in Brazil was available, from which a prior distribution was derived. 

The framework investigated here has two key advantages over the approaches based on censored likelihood discussed in Section \ref{sec:censored}. Firstly, modelling the severity of under-reporting, through the reporting probability, presents the opportunity to reduce under-reporting in the future, by informing decision-making about where additional resources for surveillance programmes would be most effective. Secondly, by modelling the under-reported counts, a more complete predictive inference on corrected counts is made available, informed by the reporting probability, the rate of the count-generating process and the recorded count. The results in Section \ref{sec:application}, for instance, provide predictions of the under-reporting rate at a micro-regional level, meaning that resources could be intelligently applied to the worst-performing areas.

\begin{appendices}
\section{Further Simulation Experiments}\label{appendix}
\subsection{Informative prior versus completely observed counts}\label{app:information}
In Section \ref{sec:background} we discussed the need to supplement the lack of information in the data, in order to distinguish between the under-reporting rate and incidence rate. This is done by either providing an informative prior distribution for $\beta_0$, the mean reporting rate at the logistic scale, or by utilising some completely reported counts, or both. In this experiment we investigate the effect of varying the strength of the informative prior and the number of completely observed counts, on predictive uncertainty.
\begin{figure}[h!]
\floatbox[{\capbeside\thisfloatsetup{capbesideposition={right,center},capbesidewidth=2.5in}}]{figure}[\FBwidth]
{\caption{Mean values of the posterior predictive log-mean squared errors for each modelling scenario.} \label{fig:lmse}}
{\includegraphics[scale=0.8]{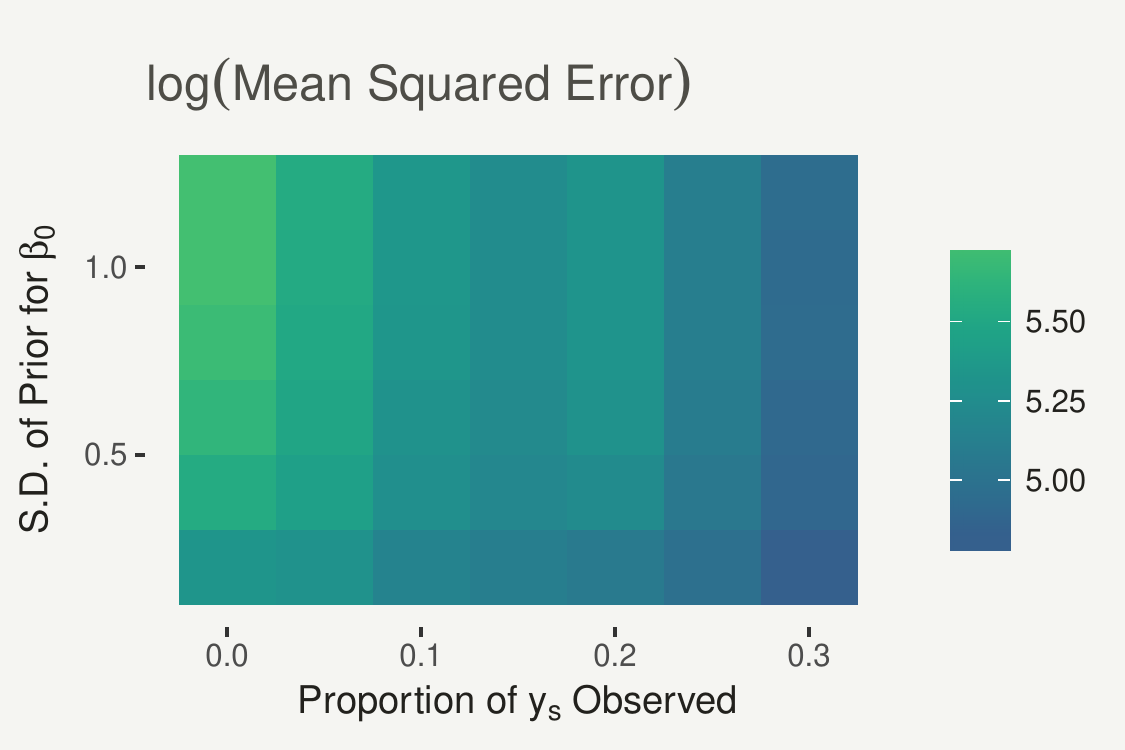}}
\end{figure}

The model was applied to simulated data, as in Section \ref{sec:sims}, using different values for the prior standard deviation, to reflect varying levels of prior certainty about the reporting rate, and including completely reported counts for varying proportions of the data. Predictive uncertainty was quantified using the logarithm of the mean squared error of $y_s$, computed for each posterior sample, which we summarise using the mean. Figure \ref{fig:lmse} shows how this uncertainty varies with prior variability in $\beta_0$ and the number of completely reported counts. The left-most column shows that predictive uncertainty decreases with increasing prior precision when there are no completely reported counts. In this case, practitioners must trade-off predictive uncertainty with the risk of systematic bias posed by specifying an overly strong prior away from the true value, seen in Section \ref{sec:sims}. While predictive uncertainty does decrease with increasing prior strength, we can also see that it decreases more substantially by increasing the proportion of counts which are known to be completely reported. This implies that the use of completely observed counts is worthwhile, if possible.

\subsection{Strength of under-reporting covariate}\label{app:strength}
\begin{figure}[h!]
\centering
\caption{Scatter plots comparing covariates $v_{s,2},...,v_{s,6}$ to the reporting probability $\pi_s$.}
\label{fig:sim_covariates}
\includegraphics[width=\linewidth]{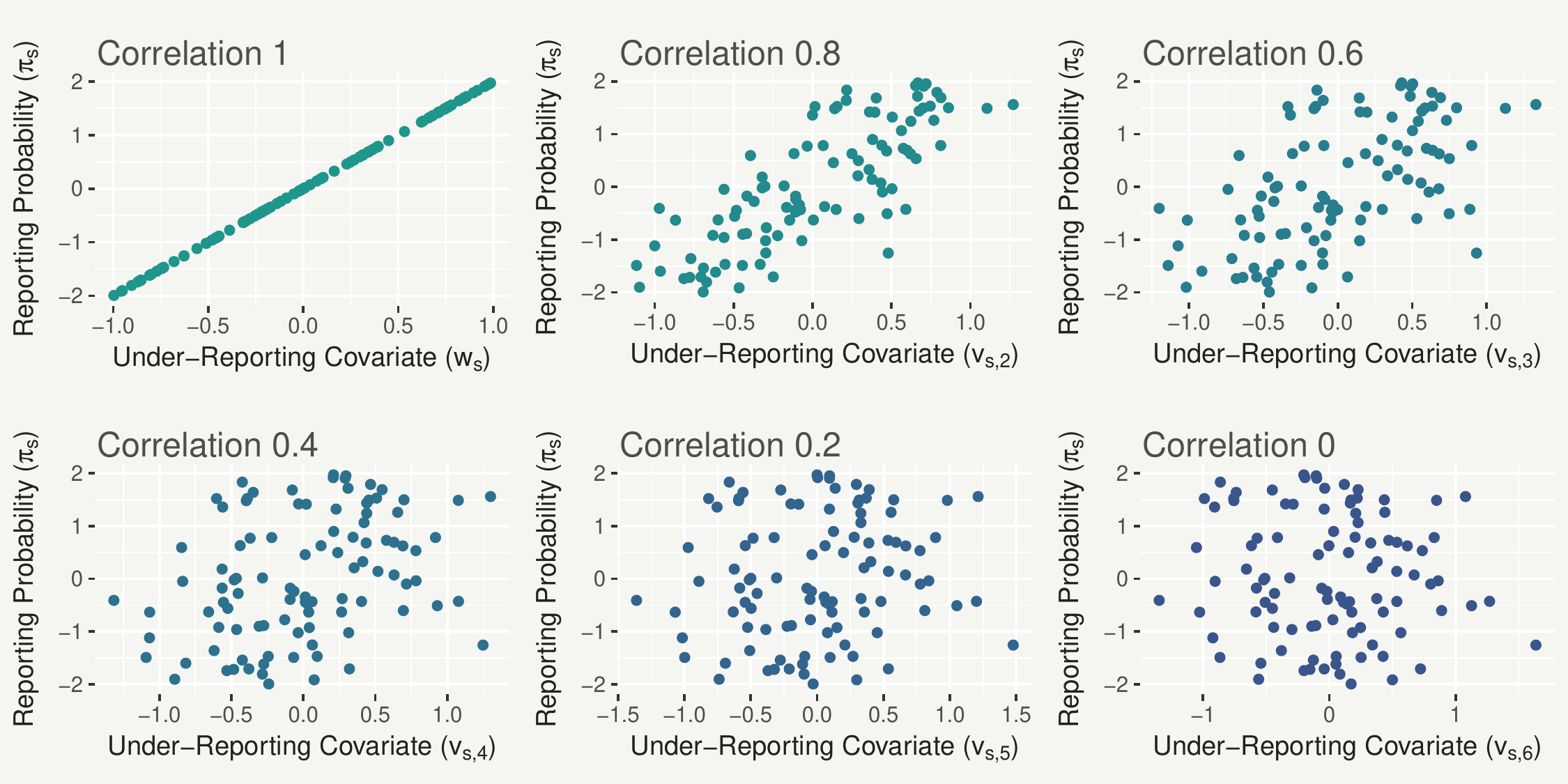}
\end{figure}

In Section \ref{sec:sims}, we varied the strength of relationship between the under-reporting covariate and the true under-reporting covariate. Figure \ref{fig:sim_covariates} shows the relationship between the different ``proxy'' covariates and the reporting probability $\pi_s$. This section presents the effect of using these proxies instead of the true under-reporting covariate $w_s$.

While the full results can be found in the Supplementary Material, the three plots in Figure \ref{fig:sim_strength} summarise the effect that varying the strength of this covariate has on the performance of the model, using locally weighted scatterplot smoothing (LOESS). The left plot shows the 95\% PI coverage. As discussed in Section \ref{sec:sims}, coverage should not decrease with covariate strength, and indeed there is very little evidence of any change. The central plot shows the mean error of $\log{(\lambda_s)}$. Again, the plot shows little evidence that this changes with covariate strength, which is reassuring as it suggests that using a weaker covariate does not necessarily introduce any systematic bias. Finally, the right plot shows a substantial effect of covariate strength on the predictive accuracy of $\log{(\lambda_s)}$, with stronger covariates translating to higher predictive accuracy, which is expected.

\begin{figure}[h!]
\centering
\caption{Scatter plots comparing the correlation of the under-reporting covariate used, from the set $v_{s,1},...,v_{s,6}$, to 95\% PI coverage for the true counts $y_s$ (left), the mean error of $\log{(\lambda_s)}$ (centre) and the square root of the mean squared error of $\log{(\lambda_s)}$ (right).}
\label{fig:sim_strength}
\includegraphics[width=\linewidth]{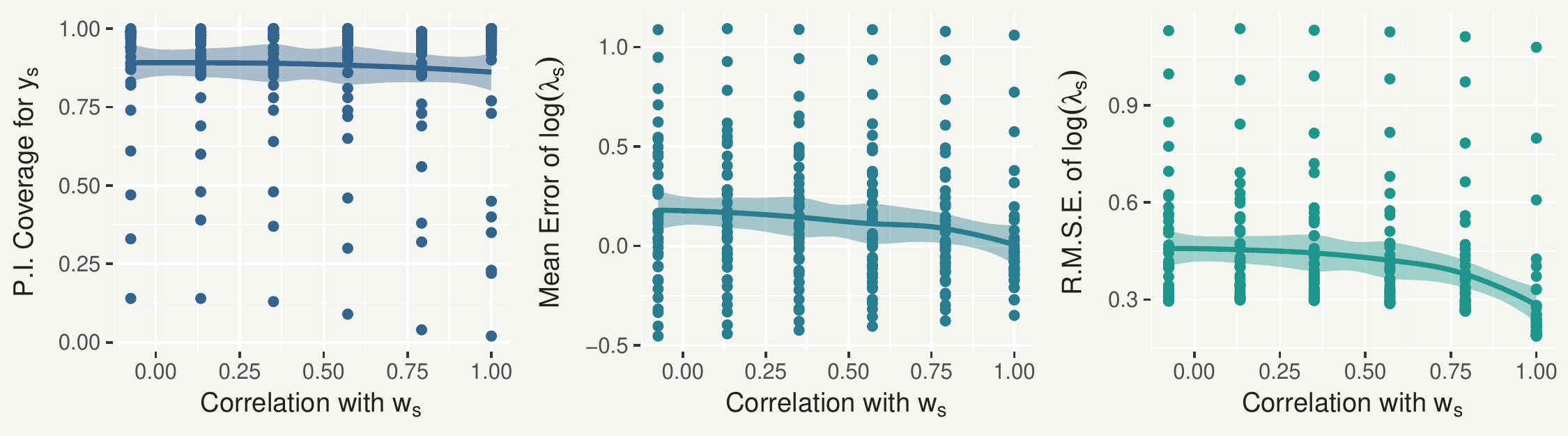}
\end{figure}

This experiment suggests that gains in predictive accuracy can be achieved by using covariates that are only proxies of the under-reporting process, compared to not including them, without necessarily introducing bias. However, this relies on those covariates being correctly identified as being related to the under-reporting mechanism. The following section illustrates the risks associated with this classification.

\subsection{Classification of covariates}\label{app:class}
In the application to TB data, the classification of covariates into those that relate to the under-reporting mechanism and those related to the true count generating process was relatively straightforward. In general, this can be more challenging and in this section we present the effects of incorrectly classifying covariates.

The experiment begins by using simulated data from the model in Section \ref{sec:sims}, with the exception of an additional unstructured random effect in the model for $\lambda$. The prior distributions are the same, with a $\mbox{N}(0,0.6^2)$ prior on $\beta_0$. In the first instance, the model is correctly informed that covariate $x_s$ belongs in the model for $\lambda_s$ and $w_s$ belongs in the model for $\pi_s$. In the second instance, these are swapped. For comparison, the model is also applied with no covariates included.
\begin{figure}[h!]
\centering
\caption{Scatter plots comparing the true simulated counts $y_s$ to the median predicted counts from the model where the covariates are classified correctly (left) and incorrectly (right), and the model where the covariates are not included (centre).}
\label{fig:sim_class}
\includegraphics[width=\linewidth]{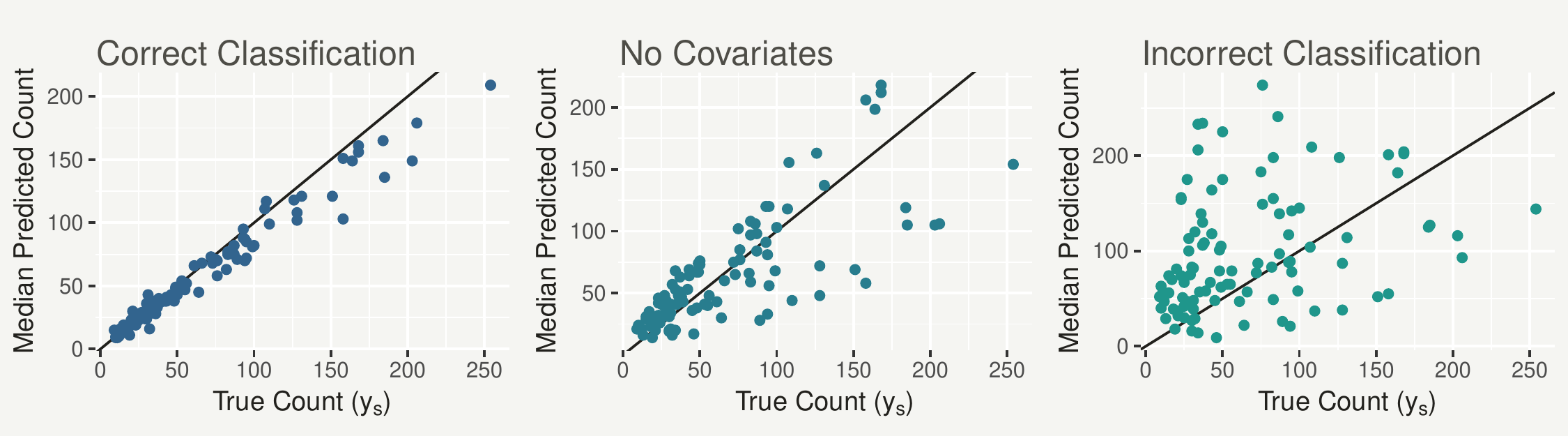}
\end{figure}

Figure \ref{fig:sim_class} shows scatter plots for each case, comparing median predicted values for $y_s$ to their corresponding true values. The left plot shows that when the covariates are correctly classified, the model is able to detect the unobserved $y_s$ values very well. When the covariates are incorrectly classified (right), the model performs very poorly. In fact, in this case the model performs even worse than a model where no covariates are included and the only random effects are relied upon to improve predictions (centre).

This experiment highlights the sensitivity of the framework to the classification of covariates, which represents an informative choice. In our view, if there is substantial doubt about whether a covariate likely relates to the under-reporting mechanism or to the true count process, it may be wiser to not include it in the model, which in this experiment results in better predictive performance.
\end{appendices}

\begin{center}
{\large\bf SUPPLEMENTARY MATERIAL}
\end{center}

All supplementary files are contained within an archive which is available to download as a single file.

\begin{description}
\item[Master:] This is the master script which the other scripts may be run from to produce all figures in the article. (R script)

\item[Simulation:] This script produces simulated data as in Section \ref{sec:sims}. (R script)

\item[Experiments:] This file reproduces the simulation experiments found in Section \ref{sec:sims} and the Appendix. (R script)

\item[Tuberculosis:] This script contains the necessary code to run the tuberculosis model  (whilst also using the Data workspace), reproducing the results found in Section \ref{sec:check} and Section \ref{sec:results}. (R script)

\item[Functions:] This script contains miscellaneous functions needed for the analysis. (R script)

\item[Data:] This file contains the data needed to execute code in the Tuberculosis script. (R workspace)

\item[Definitions:] Descriptions and sources of the variables included in the TB data. (PDF document)

\end{description}
\bibliographystyle{chicago}
\bibliography{library} 

\begin{thebibliography}{}

\bibitem[\protect\citeauthoryear{Agresti}{Agresti}{2002}]{Agresti2002}
Agresti, A. (2002).
\newblock {\em Categorical Data Analysis}.
\newblock Wiley.

\bibitem[\protect\citeauthoryear{{Atlas}}{{Atlas}}{2013}]{Atlas}
{Atlas} (2013).
\newblock The brazilian municipal human development index.
\newblock Technical report.

\bibitem[\protect\citeauthoryear{Bailey, Carvalho, Lapa, Souza, and
  Brewer}{Bailey et~al.}{2005}]{BAILEY2005335}
Bailey, T., M.~Carvalho, T.~Lapa, W.~Souza, and M.~Brewer (2005).
\newblock Modeling of under-detection of cases in disease surveillance.
\newblock {\em Annals of Epidemiology\/}~{\em 15\/}(5), 335 -- 343.

\bibitem[\protect\citeauthoryear{Besag, York, and Molli{\'e}}{Besag
  et~al.}{1991}]{Besag1991}
Besag, J., J.~York, and A.~Molli{\'e} (1991, Mar).
\newblock Bayesian image restoration, with two applications in spatial
  statistics.
\newblock {\em Annals of the Institute of Statistical Mathematics\/}~{\em
  43\/}(1), 1--20.

\bibitem[\protect\citeauthoryear{Brooks and Gelman}{Brooks and
  Gelman}{1998}]{convergence}
Brooks, S.~P. and A.~Gelman (1998).
\newblock General methods for monitoring convergence of iterative simulations.
\newblock {\em Journal of Computational and Graphical Statistics\/}~{\em
  7\/}(4), 434--455.

\bibitem[\protect\citeauthoryear{de~Valpine, Turek, Paciorek, Anderson-Bergman,
  Lang, and Bodik}{de~Valpine et~al.}{2017}]{nimble}
de~Valpine, P., D.~Turek, C.~J. Paciorek, C.~Anderson-Bergman, D.~T. Lang, and
  R.~Bodik (2017).
\newblock Programming with models: Writing statistical algorithms for general
  model structures with nimble.
\newblock {\em Journal of Computational and Graphical Statistics\/}~{\em
  26\/}(2), 403--413.

\bibitem[\protect\citeauthoryear{Dvorzak and Wagner}{Dvorzak and
  Wagner}{2016}]{Dvorzak2016}
Dvorzak, M. and H.~Wagner (2016).
\newblock Sparse bayesian modelling of underreported count data.
\newblock {\em Statistical Modelling\/}~{\em 16\/}(1), 24--46.

\bibitem[\protect\citeauthoryear{Gelman, Carlin, Stern, Dunson, Vehtari, and
  Rubin}{Gelman et~al.}{2014}]{Gelman2013}
Gelman, A., J.~Carlin, H.~Stern, D.~Dunson, A.~Vehtari, and D.~Rubin (2014,
  November).
\newblock {\em Bayesian Data Analysis, Third Edition (Chapman and {Hall/CRC}
  Texts in Statistical Science)\/} (Third ed.).
\newblock London: Chapman and Hall/CRC.

\bibitem[\protect\citeauthoryear{Greer, Stamey, and Young}{Greer
  et~al.}{2011}]{GREER}
Greer, B.~A., J.~D. Stamey, and D.~M. Young (2011).
\newblock Bayesian interval estimation for the difference of two independent
  poisson rates using data subject to under-reporting.
\newblock {\em Statistica Neerlandica\/}~{\em 65\/}(3), 259--274.

\bibitem[\protect\citeauthoryear{Kennedy and Gentle}{Kennedy and
  Gentle}{1980}]{Kennedy}
Kennedy, W. and J.~Gentle (1980).
\newblock {\em Statistical Computing}.
\newblock Marcel Dekker.

\bibitem[\protect\citeauthoryear{Moreno and Girón}{Moreno and
  Girón}{1998}]{MORENO}
Moreno, E. and J.~Girón (1998).
\newblock Estimating with incomplete count data a bayesian approach.
\newblock {\em Journal of Statistical Planning and Inference\/}~{\em 66\/}(1),
  147 -- 159.

\bibitem[\protect\citeauthoryear{Oliveira, Loschi, and Assunção}{Oliveira
  et~al.}{2017}]{Oliveira2017}
Oliveira, G.~L., R.~H. Loschi, and R.~M. Assunção (2017).
\newblock A random-censoring poisson model for underreported data.
\newblock {\em Statistics in Medicine\/}~{\em 36\/}(30), 4873--4892.

\bibitem[\protect\citeauthoryear{Papadopoulos and Silva}{Papadopoulos and
  Silva}{2008}]{Papadopoulos2008}
Papadopoulos, G. and J.~M. C.~S. Silva (2008).
\newblock Identification issues in models for underreported counts.
\newblock {\em Discussion Paper Series, Department of Economics, University of
  Essex\/}~(657).

\bibitem[\protect\citeauthoryear{Papadopoulos and Silva}{Papadopoulos and
  Silva}{2012}]{identifiability2012}
Papadopoulos, G. and J.~S. Silva (2012).
\newblock Identification issues in some double-index models for non-negative
  data.
\newblock {\em Economics Letters\/}~{\em 117\/}(1), 365 -- 367.

\bibitem[\protect\citeauthoryear{{R Core Team}}{{R Core Team}}{2018}]{R}
{R Core Team} (2018).
\newblock {\em R: A Language and Environment for Statistical Computing}.
\newblock Vienna, Austria: R Foundation for Statistical Computing.

\bibitem[\protect\citeauthoryear{Shaweno, Trauer, Denholm, and McBryde}{Shaweno
  et~al.}{2017}]{Shaweno2017}
Shaweno, D., J.~M. Trauer, J.~T. Denholm, and E.~S. McBryde (2017, Oct).
\newblock A novel bayesian geospatial method for estimating tuberculosis
  incidence reveals many missed tb cases in ethiopia.
\newblock {\em BMC Infectious Diseases\/}~{\em 17\/}(1), 662.

\bibitem[\protect\citeauthoryear{Stamey, Young, and Boese}{Stamey
  et~al.}{2006}]{STAMEY}
Stamey, J.~D., D.~M. Young, and D.~Boese (2006).
\newblock A bayesian hierarchical model for poisson rate and
  reporting-probability inference using double sampling.
\newblock {\em Australian and New Zealand Journal of Statistics\/}~{\em
  48\/}(2), 201--212.

\bibitem[\protect\citeauthoryear{Stoner}{Stoner}{2018}]{Stoner18}
Stoner, O. (2018, July).
\newblock Correcting under-reporting in historical volcano data.
\newblock {\em Proceedings of the 33rd International Workshop on Statistical
  Modelling\/}~{\em 1}, 288--292.

\bibitem[\protect\citeauthoryear{Tibbits, Groendyke, Haran, and
  Liechty}{Tibbits et~al.}{2014}]{AFSS}
Tibbits, M.~M., C.~Groendyke, M.~Haran, and J.~C. Liechty (2014).
\newblock Automated factor slice sampling.
\newblock {\em Journal of Computational and Graphical Statistics\/}~{\em
  23\/}(2), 543--563.

\bibitem[\protect\citeauthoryear{Winkelmann}{Winkelmann}{1996}]{Winkelmann1996}
Winkelmann, R. (1996, Dec).
\newblock Markov chain monte carlo analysis of underreported count data with an
  application to worker absenteeism.
\newblock {\em Empirical Economics\/}~{\em 21\/}(4), 575--587.

\bibitem[\protect\citeauthoryear{Winkelmann}{Winkelmann}{1998}]{Winkelmann1998}
Winkelmann, R. (1998).
\newblock Count data models with selectivity.
\newblock {\em Econometric Reviews\/}~{\em 17\/}(4), 339--359.

\bibitem[\protect\citeauthoryear{Winkelmann}{Winkelmann}{2008}]{Winkelmann2008}
Winkelmann, R. (2008).
\newblock {\em Econometric Analysis of Count Data\/} (5th ed.).
\newblock Springer Publishing Company, Incorporated.

\bibitem[\protect\citeauthoryear{Winkelmann and Zimmermann}{Winkelmann and
  Zimmermann}{1993}]{pogit1993}
Winkelmann, R. and K.~F. Zimmermann (1993).
\newblock Poisson-logistic regression.
\newblock {\em Discussion Papers, Department of Economics, University of
  Munich\/}~{\em 93\/}(18).

\bibitem[\protect\citeauthoryear{{World Health Organization}}{{World Health
  Organization}}{2012}]{TBInventory}
{World Health Organization} (2012).
\newblock {\em Assessing tuberculosis under-reporting through inventory
  studies}.
\newblock Geneva, Switzerland.

\bibitem[\protect\citeauthoryear{{World Health Organization}}{{World Health
  Organization}}{2017}]{TBReport}
{World Health Organization} (2017).
\newblock {\em Global Tuberculosis Report}.
\newblock Geneva, Switzerland.

\end{thebibliography}
\end{document}